\documentstyle[prd,aps]{revtex}

\newcommand{\bea}{\begin{eqnarray}}
\newcommand{\eea}{\end{eqnarray}}

\begin{document}
%%%%%%%%%%%%%%%%%%%%%%%%%%%%%%%%%%%%%%%%%%%%%%%%%%%%%%%%%%%%%%%
\draft
%  For 2 column format.
\twocolumn[\hsize\textwidth\columnwidth\hsize\csname
@twocolumnfalse\endcsname

%%%%%%%%%%%%%%%%%%%%%%%%%%%%%%%%%%%%%%%%%%%%%%%%%%%%%%%%%%%%%%%
\title{Cosmological perturbations in a generalized gravity
       including tachyonic condensation}
\author{Jai-chan Hwang${}^{(a)}$ and Hyerim Noh${}^{(b)}$ \\
        ${}^{(a)}$ {\sl Department of Astronomy and Atmospheric Sciences,
                        Kyungpook National University, Taegu, Korea} \\
        ${}^{(b)}$ {\sl Korea Astronomy Observatory, Taejon, Korea}
        }
\date{\today}
\maketitle

%%%%%%%%%%%%%%%%%%%%%%%%%%%%%%%%%%%%%%%%%%%%%%%%%%%%%%%%%%%%%%%
\begin{abstract}

We present unified ways of handling the cosmological perturbations
in a class of gravity theory covered by a general action in eq. (1).
This gravity includes our previous generalized $f(\phi,R)$ gravity 
and the gravity theory motivated by the tachyonic condensation.
We present general prescription to derive the power spectra generated 
from vacuum quantum fluctuations in the slow-roll inflation era.
An application is made to a slow-roll inflation based on the
tachyonic condensation with an exponential potential.

\end{abstract}

PACS numbers: 04.62.+v, 98.80-k, 98.80.Cq, 98.80.Hw

%%%%%%%%%%%%%%%%%%%%%%%%%%%%%%%%%%%%%%%%%%%%%%%%%%%%%%%%%%%%%%%
%  For 2 column format.
\vskip2pc]

%%%%%%%%%%%%%%%%%%%%%%%%%%%%%%%%%%%%%%%%%%%%%%%%%%%%%%%%%%%%%%%
\section{Introduction}
                                             \label{sec:Introduction}

In our present paradigm of physical cosmology, the observed large-scale 
cosmic structures and the anisotropies of the cosmic microwave 
background (CMB) are regarded as small deviations from the spatially 
homogeneous and isotropic Friedmann world model \cite{Friedmann-1922}.
In such a paradigm the structures in the large-scale limit and 
in the early stage of the evolution are assumed to be linear deviations 
from the background world model \cite{Lifshitz-1946}.
Although the observations are consistent with the perturbed Friedmann
world model, these, however, do not necessarily constrain the underlying 
gravity theory (and the matter content) to be the Einstein one.
Generalized forms of gravity appear in variety of situations involving 
the quantum aspects of the gravity theory and the low energy limits of 
the unified theories of gravity with other fundamental forces.
Thus, it is likely that the early stages of the universe
were governed by the gravity more general than Einstein one.

We have been studying the cosmological perturbations in the so called
$f(\phi,R)$ gravity theory which includes diverse generalized gravity
theories known in the literature as cases, \cite{H-GGT-1990}.
In this work, motivated by the recent interests on the action
based on the tachyonic condensation \cite{tachyon}, 
and also by a previous study in the context of ``k-inflation'' 
\cite{k-inflation}, we extend our study to a
more general form of gravity presented in eq. (\ref{action}).
Section \ref{sec:Classical} presents the classical evolutions
in a unified form. 
Section \ref{sec:Slow-roll}
presents the quantum generation process and the generated
power spectra under the slow-roll assumption and others.
Section \ref{sec:Tachyon} is an application a tachyonic slow-roll inflation.
We set $c \equiv 1 \equiv \hbar$.

%%%%%%%%%%%%%%%%%%%%%%%%%%%%%%%%%%%%%%%%%%%%%%%%%%%%%%%%%%%%%%%
\section{Gravity}
                                             \label{sec:Gravity}

We consider an action
\bea
   & & S = \int d^4 x \sqrt{-g} \left[ {1 \over 2} f (R, \phi, X) + L_m \right],
   \label{action}
\eea
where $X \equiv {1 \over 2} \phi^{;c} \phi_{,c}$, and $f$ is a
general algebraic function of $R$, $\phi$ and $X$.
This action includes the following gravity theories as cases.
(1) A minimally coupled scalar field:
    $f = {1 \over 8 \pi G} R - 2 X - 2 V(\phi)$.
(2) $f(\phi, R)$ gravity:
    $f = \tilde f(\phi, R) - 2 \omega (\phi) X - 2 V(\phi)$.
(3) $p(\phi, X)$ gravity:
    $f = {1 \over 8 \pi G} R + 2 p(\phi, X) $.
(4) Tachyonic condensation:
    $f = {1 \over 8 \pi G} R - 2 V(\phi) \sqrt{1 + 2 X}$.

The gravitational field equation and the equation of motion become
\bea
   & & G_{ab} = {1 \over F} \Big[ T^{(m)}_{ab} 
       + {1 \over 2} \left( f - FR \right)
       g_{ab} + F_{,a;b} - F^{;c}_{\;\;\; c} g_{ab}
   \nonumber \\
   & & \quad
       - {1 \over 2} f_{,X} \phi_{,a} \phi_{,b} \Big]
       \equiv 8 \pi G T_{ab},
   \label{GFE} \\
   & & \left( f_{,X} \phi^{;c} \right)_{;c} = f_{,\phi},
   \label{EOM} \\
   & & T_{(m)a;b}^{\;\;\;\;\; b} = 0,
   \label{matter-conservation}
\eea
where $F \equiv f_{,R}$.
$T_{ab}$ is the effective energy-momentum tensor, and
$T^{(m)}_{ab}$ is the energy-momentum tensor of additional matters.

%%%%%%%%%%%%%%%%%%%%%%%%%%%%%%%%%%%%%%%%%%%%%%%%%%%%%%%%%%%%%%%
\section{Classical perturbations}
                                              \label{sec:Classical}

We consider the Friedmann background with the scalar- and the tensor-type
perturbations.
Our metric convention follows Bardeen's \cite{Bardeen-1988}
\bea
   & & d s^2 = - a^2 \left( 1 + 2 \alpha \right) d \eta^2
       - 2 a^2 \beta_{,\alpha} d \eta d x^\alpha
   \nonumber \\
   & & \quad
       + a^2 \Big[ g^{(3)}_{\alpha\beta} \left( 1 + 2 \varphi \right)
       + 2 \gamma_{,\alpha|\beta} 
       + 2 C_{\alpha\beta} \Big] d x^\alpha d x^\beta.
   \label{metric}
\eea
The energy-momentum tensor is decomposed as
\bea
   & & T^0_0 = - \left( \bar \mu + \delta \mu \right), \quad
       T^0_\alpha = - \left( \mu + p \right) v_{,\alpha}/k,
   \nonumber \\
   & & T^\alpha_\beta = \left( \bar p + \delta p \right) \delta^\alpha_\beta
       + \Big( {1 \over k^2} \nabla^{(3)\alpha} \nabla^{(3)}_\beta
       + {1 \over 3} \delta^\alpha_\beta \Big) \pi^{(s)}
       + \pi^{\alpha}_{\beta}.
   \label{Tab}
\eea
A vertical bar $|$ and $\nabla^{(3)}_\alpha$ are the covariant derivatives
based on $g^{(3)}_{\alpha\beta}$.

To the background order, eq. (\ref{GFE}) gives
\bea
   & & H^2 = {8 \pi G \over 3} \mu - {K \over a^2}, \quad
       \dot H = - 4 \pi G \left( \mu + p \right) + {K \over a^2},
   \label{BG-eqs}
\eea
where $H \equiv {\dot a \over a}$ and an overdot denotes a time 
derivative based on $t$ with $dt \equiv a d \eta$.
We also have $R = 6 ( 2 H^2 + \dot H + {K \over a^2} )$.
The effective fluid quantities are
\bea
   & & 8 \pi G \mu = {1 \over F} \left[ \mu^{(m)}
       - {1 \over 2} ( f - F R ) - {1 \over 2} f_{,X} \dot \phi^2
       - 3 H \dot F \right],
   \nonumber \\
   & & 8 \pi G p = {1 \over F} \left[ p^{(m)}
       + {1 \over 2} ( f - F R ) + \ddot F + 2 H \dot F \right],
   \label{BG-fluids}
\eea
where we have $X = - {1 \over 2} \dot \phi^2$.
To the background order eq. (\ref{EOM}) gives
\bea
   & & {1 \over a^3} \left( a^3 f_{,X} \dot \phi \right)^\cdot
       + f_{,\phi} = 0.
   \label{BG-EOM}
\eea

Perturbed set of equations can be derived similarly.
The perturbed set of equations in Einstein gravity
based on our convention in eqs. (\ref{metric},\ref{Tab}) is presented in 
\cite{Bardeen-1988}\footnote{
     See eqs. (43)-(50) in \cite{HN-GGT-1996}.
     We have $\epsilon \equiv \delta \mu$, $\pi = \delta p$,
     $\Psi \equiv - {a\over k} ( \mu + p) v$, and
     $\sigma = {a^2 \over k^2} \pi^{(s)}$.
     }.
These equations are valid even in our gravity theory if we 
re-interprete the fluid quantities as the effective ones.
The perturbed order effective fluid quantities can be easily read
by comparing eq. (\ref{Tab}) with eq. (\ref{GFE}).

%-----------------------------------------------------------
{}For the scalar-type perturbation we ignore the presence of
additional fluid, thus $T^{(m)}_{ab} = 0$.
In the following we consider two general situations:
     (i) $F=F(\phi)$ and $K=0$, and 
     (ii) $F = {1 \over 8 \pi G}$ but general $K$.
We introduce the Field-Shepley combination 
\cite{Field-Shepley-1968}\footnote{ 
     See the paragraph containing eq. (36) in \cite{HN-Fluids-2002}.
     }
\bea
   & & \Phi \equiv \varphi_{\delta \phi}
       - {K / a^2 \over 4 \pi G ( \mu + p )} \varphi_\chi,
\eea
where 
\bea
   & & \varphi_{\delta \phi} \equiv \varphi - (H / \dot \phi) \delta \phi, 
       \quad
       \varphi_\chi \equiv \varphi - H \chi,
\eea
are gauge-invariant combinations\footnote{
     $\varphi_{\delta \phi}$ is the same $\varphi$ in the 
     uniform-field gauge ($\delta \phi \equiv 0$) \cite{Lukash-Mukhanov}.
     $\varphi_\chi$ is the same as $\varphi$ in the zero-shear gauge 
     ($\chi \equiv 0$) \cite{Harrison-1967}, and is the same as $\Phi_H$ 
     which is often called the Bardeen potential \cite{Bardeen-1980}.
     };
$\chi \equiv a (\beta + a \dot \gamma)$ is a spatially gauge-invariant
combination \cite{Bardeen-1988}.

(i) In the first case, perturbed parts of eq. (\ref{GFE}) can be combined 
to give a closed form of second-order differential equation for 
$\varphi_{\delta \phi}$\footnote{The procedure is exactly the same
     as the one used to derive eq. (66) in \cite{HN-GGT-1996}.
     }
\bea
   & & {1 \over a^3 Q} \left( a^3 Q \dot \varphi_{\delta \phi} \right)^\cdot
       + c_A^2 {k^2 \over a^2} \varphi_{\delta \phi} = 0,
   \label{varphi_delta-phi-eq} \\
   & & Q \equiv { {3 \dot F^2 \over 2 F} + f_{,X} X + 2 f_{,XX} X^2
       \over \left( H + {\dot F \over 2F} \right)^2 }
       \equiv {\dot \phi^2 \over H^2} Z, 
   \nonumber \\
   & & c_A^2 \equiv \left( 1 + {2 f_{,XX} X^2 \over
       {3 \dot F^2 \over 2 F} + f_{,X} X} \right)^{-1}.
   \label{Q-c_A}
\eea
{}For $f_{,X} = - 2 \omega (\phi)$ we recover the result derived in the
$f(\phi,R)$ gravity theory \cite{HN-GGT-1996}.

(ii) In the second case, perturbed parts of eq. (\ref{GFE}) can be combined to 
give\footnote{The procedure is exactly the same
     as the one used to derive eqs. (32,33) in \cite{HN-Fluids-2002}.
     }
\bea
   & & \Phi = {H^2 \over 4 \pi G ( \mu + p ) a}
       \left( {a \over H} \varphi_\chi \right)^\cdot,
   \label{Phi-1} \\
   & & \dot \Phi = - {H c_A^2 k^2 \over 4 \pi G ( \mu + p ) a^2}
       \varphi_\chi,
   \label{Phi-2}
\eea
where $\mu + p = - {1 \over 2} f_{,X} \dot \phi^2 = f_{,X} X$, and
\bea
   & & c_A^2 \equiv c_X^2 - {p_{,\phi} - c_X^2 \mu_{,\phi} \over \mu + p}
       \dot \phi {K \over k^2}, 
   \nonumber \\
   & & c_X^2 \equiv {p_{,X} \over \mu_{,X}}
       = { f_{,X} \over f_{,X} + 2 f_{,XX} X }. 
\eea
Equations (\ref{Phi-1},\ref{Phi-2}) were derived by Garriga and Mukhanov;
see eqs. (21,22) in \cite{Garriga-Mukhanov-1999}.
Equations (\ref{Phi-1},\ref{Phi-2}) can be combined to give
\bea
   & & {1 \over a^3 Q} \left( a^3 Q \dot \Phi \right)^\cdot
       + c_A^2 {k^2 \over a^2} \Phi = 0, \quad
       Q \equiv {\mu + p \over c_A^2 H^2},
   \label{Phi-eq} \\
   & & {\mu + p \over H} \left[ {H^2 \over ( \mu + p ) a}
       \left( {a \over H} \varphi_\chi \right)^\cdot \right]^\cdot
       + c_A^2 {k^2 \over a^2} \varphi_\chi = 0.
   \label{varphi_chi-eq} 
\eea
Using
\bea
   & & v \equiv z \Phi, \quad
       u \equiv {\varphi_\chi \over \sqrt{\mu + p}}, \quad
       z \equiv a \sqrt{Q} %= {a \sqrt{\mu + p} \over c_A H}
       \equiv {1 \over c_A} \tilde z,
\eea
eqs. (\ref{Phi-eq},\ref{varphi_chi-eq}) become
the well known equations \cite{Field-Shepley-1968,Lukash-Mukhanov}
\bea
   & & v^{\prime\prime} + \left( c_A^2 k^2 - {z^{\prime\prime} \over z}
       \right) v = 0,
   \label{v-eq} \\
   & & u^{\prime\prime} + \left( c_A^2 k^2
       - {(1/\tilde z)^{\prime\prime} \over 1/\tilde z} \right) u = 0,
   \label{u-eq}
\eea
where a prime indicates a time derivative based on $\eta$.
Equation (\ref{v-eq}) is valid for the first case in 
eq. (\ref{varphi_delta-phi-eq}) as well.

In the large-scale limit, with 
$z^{\prime\prime} / z \gg c_A^2 k^2$ and 
$\tilde z (1/\tilde z)^{\prime\prime} \gg c_A^2 k^2$, we have
exact solutions
\bea
   & & \Phi = C ({\bf x}) - D ({\bf x}) \int_0^t {dt \over a z^2},
   \label{Phi-LS-sol} \\
   & & \varphi_\chi = 4 \pi G {H \over a} \left[ C ({\bf x})
       \int_0^t {\tilde z^2 \over a} dt + {1 \over k^2} D ({\bf x}) \right].
   \label{varphi_chi-LS-sol}
\eea
Ignoring the transient solution (which is the $D$-mode in expanding phases)
we have a temporally conserved behavior for $\Phi$
\bea
   & & \Phi ({\bf x}, t) = C ({\bf x}).
   \label{conservation}
\eea

%-----------------------------------------------------------
{}For the tensor-type perturbation, for the {\it general} action
in eq. (\ref{action}), we have
\bea
   & & \ddot C^\alpha_\beta + \left( 3 H + {\dot F \over F} \right) 
       \dot C^\alpha_\beta + {k^2 + 2 K \over a^2} C^\alpha_\beta
       = {1 \over F} \pi^{(m)\alpha}_{\;\;\;\;\;\; \beta},
   \label{GW}
\eea
which is the same as eq. (111) in \cite{HN-GGT-1996}
based on $f(\phi,R)$ gravity.
Thus, the presence of general algebraic complication of $X$ in 
eq. (\ref{action}) has {\it no effect} on the tensor-type perturbation. 
Also, eq. (\ref{GW}) can be written as in eqs. (\ref{Phi-eq},\ref{v-eq}).
In such cases we have $\Phi = C^\alpha_\beta$,
$Q = F \equiv Z/(8 \pi G)$, $c_A^2 = 1$, thus $z \equiv a \sqrt{F}$, and
eqs. (\ref{Phi-LS-sol},\ref{conservation}) also remain valid.

%-----------------------------------------------------------
The vector-type perturbation of additionally present fluid(s)
is described by eq. (\ref{matter-conservation}) which is {\it not} 
affected by the generalized nature of the gravity theory in eq. (\ref{action}).

%%%%%%%%%%%%%%%%%%%%%%%%%%%%%%%%%%%%%%%%%%%%%%%%%%%%%%%%%%%%%%%
\section{Slow-roll inflation}
                                              \label{sec:Slow-roll}

As in \cite{NH-Unified-2001} the quantum generation 
process can be presented in a unified form.
{}From eq. (\ref{Phi-eq}) we can construct the perturbed action
\cite{Lukash-Mukhanov}
\bea
   & & \delta^2 S = {1 \over 2} \int a^3 Q \left( \dot \Phi^2
       - c_A^2 {1 \over a^2} \Phi^{|\gamma} \Phi_{,\gamma}
       \right) dt d^3 x,
   \label{perturbed-action}
\eea
which is valid for both the scalar-type and tensor-type perturbations
in a unified form. 
The rest of the canonical quantization process is straightforward, see
\cite{NH-Unified-2001}.
Under an {\it ansatz}
\bea
   & & z^{\prime\prime}/z = n/\eta^2, \quad
       c_A^2 = {\rm constant},
   \label{ansatz}
\eea
where $n = n_s, \; n_t$ for the two perturbation types, 
the mode function has an exact solution in terms of the Hankel functions,
see eq. (24) in \cite{NH-Unified-2001}.
The power spectrum based on the vacuum expectation value of $\hat \Phi$ 
can be constructed as in eq. (26) of \cite{NH-Unified-2001}, and
in the large-scale limit we have\footnote{
     {}For $\nu = 0$ we have an additional $2 \ln{(c_A k |\eta|)}$ factor.
     {}For the gravitational we should consider additional $\sqrt{2}$ factor 
     \cite{Hwang-GW-1998}.
     }
\bea
   & & {\cal P}_{\hat \Phi}^{1/2} \Big|_{LS}
       = {H \over 2 \pi} {1 \over a H |\eta|}
       {\Gamma(\nu) \over \Gamma(3/2)}
       \left( {k |\eta| \over 2} \right)^{3/2 - \nu}
       {1 \over c_A^\nu \sqrt{Q}},
   \label{P-LS}
\eea
where $\nu \equiv \sqrt{n+1/4}$.
We can read the spectral indices
\bea
   & & n_S - 1 = 3 - \sqrt{4 n_s + 1}, \quad
       n_T = 3 - \sqrt{4 n_t + 1}.
   \label{n-def}
\eea

We introduce the slow-roll parameters \cite{HN-GGT-1996}
\bea
   & & \epsilon_1 \equiv {\dot H \over H^2}, \quad
       \epsilon_2 \equiv {\ddot \phi \over H \dot \phi}, \quad
       \epsilon_3 \equiv {1 \over 2} {\dot F \over H F}, \quad
       \epsilon_4 \equiv {1 \over 2} {\dot E \over H E}, 
   \nonumber \\
   & & E \equiv F \left( {3 \dot F^2 \over 2 \dot \phi^2 F} 
       - {1 \over 2} f_{,X} - f_{,XX} X \right).
   \label{slow-roll}
\eea
Compared with the Einstein gravity in \cite{Stewart-Lyth-1993}
we have two additional parameters
$\epsilon_3$ and $\epsilon_4$ for the scalar-type perturbation which
reflect the effects of additional parameters $F (\equiv f_{,R})$ and $f_{,X}$
in our generalized gravity;
for the tensor-type perturbation we have only one additional parameter
$\epsilon_3$ from $F$.
Compared with \cite{HN-GGT-1996} the only difference occurs in our definition
of $E$ which includes the $f(\phi,R)$ gravity in \cite{HN-GGT-1996} as a case.
Using our present definition of $\epsilon_i$'s our unified analyses
made in eqs. (30-32) of \cite{NH-Unified-2001} remain valid.

To the first-order in the slow-roll parameters, i.e., {\it assuming} 
\bea
   & & \dot \epsilon_i = 0, \quad
       |\epsilon_i| \ll 1,
   \label{slow-roll-condition}
\eea
we can derive
\bea
   & & {\cal P}_{\hat \varphi_{\delta \phi}}^{1/2} \Big|_{LS}
       = {H \over |\dot \phi|} {\cal P}_{\delta \hat \phi_\varphi}^{1/2}
       \Big|_{LS}
       = {H^2 \over 2 \pi |\dot \phi|} {1 \over \sqrt{Z_s}}
       \Big\{ 1 + \epsilon_1
   \nonumber \\
   & & \quad
       + \big[ \gamma_1 + \ln{(k |\eta|)} \big]
       ( 2 \epsilon_1 - \epsilon_2 + \epsilon_3 - \epsilon_4 ) \Big\}
       c_A^{-\nu_s},
   \label{P-SR-scalar} \\
   & & {\cal P}_{\hat C^\alpha_\beta}^{1/2} \Big|_{LS}
       = \sqrt{16 \pi G} {H \over 2 \pi} {1 \over \sqrt{Z_t}}
       \Big\{ 1 + \epsilon_1
   \nonumber \\
   & & \quad
       + \big[ \gamma_1 + \ln{(k |\eta|)} \big]
       ( \epsilon_1 - \epsilon_3 ) \Big\},
   \label{P-SR-tensor}
\eea
where $\gamma_1 \equiv \gamma_E + \ln{2} - 2 = - 0.7296 \dots$,
with $\gamma_E$ the Euler constant.
We have 
\bea
   & & Z_s = { E/F \over (1 + \epsilon_3)^2}, \quad
       Z_t = 8 \pi G F,
   \label{Z}
\eea
where $Z$'s become unity in Einstein gravity.
Thus, besides $\epsilon_1$,
the scalar-type perturbation is affected by
$\epsilon_2$, $\epsilon_3$ and $\epsilon_4$
(thus, $f_{,\phi}$, $F$ and $f_{,X}$), whereas the tensor-type perturbation
is affected by $\epsilon_3$ (thus, $F$) only.
The spectral indices of the scalar and tensor-type perturbations
in eq. (\ref{n-def}) become
\bea
   & & n_S - 1 = 2 ( 2 \epsilon_1 - \epsilon_2 + \epsilon_3 - \epsilon_4 ),
       \quad
       n_T = 2 ( \epsilon_1 - \epsilon_3 ).
   \label{indices}
\eea

{}For the scale independent Harrison-Zel'dovich 
($n_S -1 \simeq 0 \simeq n_T$) spectra \cite{HZ}
the CMB quadrupole anisotropy becomes
\bea
   & & \langle a_2^2 \rangle
       = \langle a_2^2 \rangle_S + \langle a_2^2 \rangle_T
       = {\pi \over 75} {\cal P}_{\varphi_{\delta \phi}}
       + 7.74 {1 \over 5} {3 \over 32} {\cal P}_{C_{\alpha\beta}},
   \label{a_2}
\eea
which is valid for $K = 0 = \Lambda$.
The four-year {\it COBE}-DMR data give
$\langle a_2^2 \rangle \simeq 1.1 \times 10^{-10}$, \cite{COBE}.
{}From eqs. (\ref{a_2},\ref{P-SR-scalar},\ref{P-SR-tensor})
the ratio between two types of perturbations
$r_2 \equiv {\langle a_2^2 \rangle_T / \langle a_2^2 \rangle_S}$ becomes
\bea
   r_2
   &=& 13.8 \times 4 \pi G {\dot \phi^2 \over H^2} \left| {Z_s \over Z_t}
       \right| c_A^{2\nu_s}
   \nonumber \\
   &=& 13.8 {1 \over ( 1 + \epsilon_3 )^2}
       \left| (\epsilon_1 - \epsilon_3) ( 1 + \epsilon_3)
       + {\dot \epsilon_3 \over H} \right| c_A^{2-n_S}
   \nonumber \\
   &\simeq& 13.8 \left| \epsilon_1 - \epsilon_3 \right| c_A
   \nonumber \\
   &\simeq& 6.92 |n_T| c_A,
   \label{ratio}
\eea
where in the last two steps we used the slow-roll conditions
in eq. (\ref{slow-roll-condition}).
In the limit of Einstein gravity we have $r_2 = - 13.8 \epsilon_1
= - 6.92 n_T$ which is independent of $V$ and is known as 
a consistency relation.
The $c_A$ factor difference from the Einstein gravity for
$p(\phi, X)$ gravity was noticed in \cite{Garriga-Mukhanov-1999}.
{}For the $f(\phi, R)$ gravity we have $c_A^2 = 1$.

%%%%%%%%%%%%%%%%%%%%%%%%%%%%%%%%%%%%%%%%%%%%%%%%%%%%%%%%%%%%%%%
\section{Tachyonic condensation}
                                           \label{sec:Tachyon}

The recently popular tachyonic condensation is a case of our gravity 
with a form $f = {1 \over 8 \pi G} R - 2 V \sqrt{1 + 2X}$:
if based on the string theory, we should regard the field in this action 
as being written in the unit where the string theory is relevant.
We have
\bea
   & & Q = {\dot \phi^2 \over H^2} {V \over (1 - \dot \phi^2)^{3/2}}, \quad
       c_A^2 = 1 - \dot \phi^2.
\eea
Equations (\ref{v-eq}) and (\ref{u-eq}) in this case were derived 
in eq. (17) of \cite{Frolov-etal-2002} and
in eq. (44) of \cite{Shiu-Wasserman-2002}, respectively.
We have $\epsilon_3 = 0$ and $E = {V \over 8 \pi G}(1 - \dot \phi^2)^{-3/2}$.

{\it Assuming} a set of {\it slow-roll} conditions
$\ddot \phi \ll 3 H \dot \phi$ and $\dot \phi^2 \ll 1$,
and under an {\it ansatz} $V \equiv V_0 e^{-\alpha \phi}$
\cite{Sen-2002}, from
eqs. (\ref{BG-eqs}-\ref{BG-EOM}) for $K = 0$ we have \cite{Sami-etal-2002}
\bea
   & & \phi = - {2 \over \alpha} 
       \ln{\left(C - {\sqrt{3} \alpha^2 M_{pl} \over 6 \sqrt{V_0} } t
       \right)}, \quad
       a \propto e^{{C \sqrt{V_0} \over \sqrt{3} M_{pl}} t
       - {\alpha^2 \over 12} t^2},
\eea
where $M_{pl}^2 \equiv 1/(8 \pi G)$.
If we set $t_i = 0$, we have $C = e^{-(\alpha/2) \phi_i}$ and $V_i = V_0 C^2$.
{}For $t \simeq t_i$ we have an accelerated expansion stage.
In such a situation we have the slow-roll conditions in 
eq. (\ref{slow-roll-condition}) are well met, with the result:
\bea
   & & \epsilon_1 = - \epsilon_2 = \epsilon_4
       = - {\alpha^2 M_{pl}^2 \over 2 V_i}, \quad
       \epsilon_3 = 0.
\eea
Thus, eq. (\ref{indices}) gives
\bea
   & & n_S -1 = 4 \epsilon_1, \quad
       n_T = 2 \epsilon_1,
\eea
and eqs. (\ref{P-SR-scalar}-\ref{Z},\ref{ratio}) reduce to
\bea
   & & {\cal P}^{1/2}_{\varphi_{\delta \phi}}
       \simeq {H^2 \over 2 \pi |\dot \phi|} {1 \over \sqrt{V}}
       \simeq {1 \over 2 \sqrt{3} \pi} {V_i \over \alpha M_{pl}^3},
   \\
   & & {\cal P}^{1/2}_{C_{\alpha\beta}}
       \simeq \sqrt{16 \pi G} {H \over 2 \pi}
       \simeq{1 \over \sqrt{6} \pi} {\sqrt{V_i} \over M_{pl}^2},
   \\
   & & r_2 = 6.92|n_T|.
\eea
Therefore, if the seed structures were generated from the vacuum quantum
fluctuation under such a slow-roll phase, the final spectra show that:
(1) the spectra are nearly scale-invariant Harrison-Zel'dovich type,
(2) the consistency relation is met, 
(3) the graviational wave is suppressed, and 
(4) the CMB quadrupole requires
\bea
   & & \langle a_2^2 \rangle \simeq {1 \over 75 \times 12 \pi}
       {V_i^2 \over \alpha^2 M_{pl}^6} \simeq 1.1 \times 10^{-10}.
\eea

We have assumed that, firstly, the seed fluctuations were generated during
the slow-roll inflation stage supported by the tachyonic condensation,
and secondly, the tachyonic gravity stage was switched successfully  
to an ordinary big-bang stage while the fluctuations stay in the
large-scale limit (see \cite{Kofman-Linde-2002} for the reheating problem); 
in such a case the relatively growing $C$-mode
fluctuation in eq. (\ref{Phi-LS-sol}) survives as the same $C$-mode 
of the curvature fluctuation $\Phi$ 
now supported by the Einstein gravity with ordinary matter. 
We have derived these results directly based on the generalized form
of gravity theory whereas the previous analyses 
\cite{Fairbairn-Tytgat-2002,Sami-etal-2002} were based on known 
formulation in Einstein gravity by using some field redefinition.

%%%%%%%%%%%%%%%%%%%%%%%%%%%%%%%%%%%%%%%%%%%%%%%%%%%%%%%%%%%%%%%
\section{Discussions}
                                             \label{sec:Discussion}

We have presented unified ways of handling the cosmological perturbations
in a class of gravity theory covered by an action in eq. (1).
Section \ref{sec:Classical} presents the classical evolutions
in a unified form, and eqs. (\ref{P-LS},\ref{n-def}) show the
generated seed fluctuations of the quantum origin under an 
assumption in eq. (\ref{ansatz}).
The rest of section \ref{sec:Slow-roll} presents the general prescription 
to derive the power spectra generated under the slow-roll assumption,
and section \ref{sec:Tachyon} is an application to a tachyonic 
slow-roll inflation.

We note that even in the gravity with {\it additional} stringy correction terms
\bea
   & & \xi(\phi) [ c_1 R_{GB}^2 + c_2 G^{ab} \phi_{;a} \phi_{;b}
       + c_3 \Box \phi \phi^{;a} \phi_{;a}
       + c_4 (\phi^{;a} \phi_{;a})^2 ],
   \nonumber \\
   & & g(\phi) R \tilde R,
\eea
in the Lagrangian, where
$R_{GB}^2 \equiv R^{abcd} R_{abcd} - 4 R^{ab} R_{ab} + R^2$ and
$R \tilde R \equiv \eta^{abcd} R_{ab}^{\;\;\;\;ef} R_{cdef}$,
we still have eqs. (\ref{varphi_delta-phi-eq},\ref{perturbed-action}) 
with more complicated $Q$ and $c_A^2$, \cite{Stringy}.
Thus, the rest of the analyses made above can be applied similarly as well, 
\cite{Stringy}.
Similar unified formulation also exists in the fluid context, 
\cite{Lukash-Mukhanov,NH-Unified-2001}.
We also have studied situation with $R^{ab} R_{ab}$ term in the action
\cite{Rab}, in which case the gravity becomes a fourth-order theory.

We would like to emphasize that our gravity theory in
eq. (\ref{action}) covers many of the modified gravity theory, and
our assumption in eq. (\ref{ansatz}) is satisfied by
most of the expansion stages (including diverse class of inflation
scenarios available in analytic forms) considered in the literature,
and we hope our slow-roll conditions in eq. (\ref{slow-roll-condition})
cover most of the specific slow-roll conditions in the 
inflation theories based on specific modified gravity theories.
We emphasize, however, that the classical evolutions studied
in section \ref{sec:Classical} are valid for the general
cosmological situations governed by our action in eq. (\ref{action}).

%%%%%%%%%%%%%%%%%%%%%%%%%%%%%%%%%%%%%%%%%%%%%%%%%%%%%%%%%%%%%%%
\vskip .5cm
\subsection*{Acknowledgments}

We thank Hongsu Kim and Hyung Doo Kim for useful discussions.
HN was supported by grant No. R04-2000-000-00008-0 from the
Basic Research Program of the Korea Science and Engineering Foundation.
JH was supported by Korea Research Foundation grants (KRF-2001-041-D00269).

%%%%%%%%%%%%%%%%%%%%%%%%%%%%%%%%%%%%%%%%%%%%%%%%%%%%%%%%%%%%%%

%%%%%%%%%%%%%%%%%%%%%%%%%%%%%%%%%%%%%%%%%%%%%%%%%%%%%%%%%%%%%%

\begin{thebibliography}{99}
\bibitem{Friedmann-1922}
         A. Friedmann, Zeitschrift f\"ur Physik {\bf 10}, 377 (1922),
            translated in: {\it Cosmological constants, papers in
            modern cosmology}, eds., J. Bernstein and F. Feinberg,
            (Columbia Univ. Press, New York, 1986), 49p.
\bibitem{Lifshitz-1946}
         E. M. Lifshitz, J. Phys. (USSR) {\bf 10}, 116 (1946).
\bibitem{H-GGT-1990}
         J. Hwang, Class. Quant. Grav. {\bf 7}, 1613 (1990);
         J. Hwang, and H. Noh, Phys. Rev. D {\bf 65}, 023512 (2002), 
            (astro-ph/0102005).
\bibitem{tachyon}
         A. Mazumdar, S. Panda, and A. P\'erez-Lorenzana, hep-ph/0107058;
         G. W. Gibbons, hep-th/0204008;
         A. Feinstein, hep-th/0204140;
         T. Padmanabhan, hep-th/0204150;
         D. Choudhury, D. Ghoshal, C. P. Jatkar, and S. Panda, hep-th/0204204;
\bibitem{k-inflation}
         C. Armend\'ariz-Pic\'on, T. Damour, and V. F. Mukhanov, 
            Phys. Lett. B {\bf 458}, 209 (1999),
            (hep-th/9904075).
\bibitem{Bardeen-1988}
         J. M. Bardeen, in: {\it Cosmology and particle physics},
               eds., L. Fang and A. Zee, (Gordon and Breach, London, 1988), 1;
         J. Hwang, Astrophys. J. {\bf 375}, 443 (1991).
\bibitem{HN-GGT-1996}
         J. Hwang, and H. Noh, Phys. Rev. D {\bf 54}, 1460 (1996).
\bibitem{Field-Shepley-1968}
         G. B. Field, and L. C. Shepley, Astrophys. Space. Sci.
                {\bf 1}, 309 (1968);
         G. V. Chibisov, and V. F. Mukhanov, Mon. Not. R. Astron. Soc.
                   {\bf 200}, 535 (1982);
         J. Hwang, and E. T. Vishniac, Astrophys. J., {\bf 353}, 1 (1990);
         V. F. Mukhanov, H. A. Feldman and R. H. Brandenberger,
               Phys. Rep. {\bf 215}, 203 (1992).
\bibitem{HN-Fluids-2002}
         J. Hwang, and H. Noh, Class. Quant. Grav. {\bf 19}, 527 (2002),
            (astro-ph/0103244).
\bibitem{Lukash-Mukhanov}
         V. N. Lukash, Sov. Phys. JETP {\bf 52}, 807 (1980);
               Sov. Phys. JETP Lett. {\bf 31}, 596 (1980);
         V. F. Mukhanov, Sov. Phys. JETP Lett. {\bf 41}, 493 (1985);
               Sov. Phys. JETP {\bf 68}, 1297 (1988).
\bibitem{Harrison-1967}
         E. R. Harrison, Rev. Mod. Phys. {\bf 39}, 862 (1967).
\bibitem{Bardeen-1980}
         J. M. Bardeen, Phys. Rev. D {\bf 22}, 1882 (1980).
\bibitem{Garriga-Mukhanov-1999}
         J. Garriga, and V. F. Mukhanov, Phys. Lett. B {\bf 458}, 219 (1999),
            (hep-th/9904176).
\bibitem{NH-Unified-2001}
         H. Noh, and J. Hwang, Phys. Lett. B, {\bf 515}, 231, (2001),
            (astro-ph/0107069).
\bibitem{Hwang-GW-1998}
         J. Hwang, Class. Quant. Grav. {\bf 15}, 1401 (1998).
\bibitem{Stewart-Lyth-1993}
         E. D. Stewart and D. H. Lyth, Phys. Lett. B {\bf 302}, 171 (1993),
            (gr-qc/9302019).
\bibitem{HZ}
         E. R. Harrison, Phys. Rev. D {\bf 1}, 2726 (1970);
         Ya. B. Zeldovich, Mon. Not. R. Astron. Soc. {\bf 160}, 1p (1972).
\bibitem{COBE}
         C. L. Bennett, {\it et. al.}, Astrophys. J. {\bf 464}, L1 (1996),
               (astro-ph/9601067).
\bibitem{Frolov-etal-2002}
         A. Frolov, L. Kofman, and A. Starobinsky, hep-th/0204187.
\bibitem{Shiu-Wasserman-2002}
         G. Shiu, and I. Wasserman, hep-th/0205003.
\bibitem{Sen-2002}
         A. Sen, hep-th/0204143.
\bibitem{Sami-etal-2002}
         M. Sami, P. Chingangbam, and T. Qureshi, hep-th/0205179.
\bibitem{Kofman-Linde-2002}
         L. Kofman, and A. Linde, hep-th/0205121.
\bibitem{Fairbairn-Tytgat-2002}
         M. Fairbairn, and M. H. G. Tytgat, hep-th/0204070.
\bibitem{Stringy}
         J. Hwang and H. Noh, Phys. Rev. D {\bf 61}, 043511 (2000),
            (astro-ph/9909480);
         K. Choi, J. Hwang, and K. W. Hwang, {\it ib. id.} {\bf 61}, 084026
            (2000), (hep-ph/9907244);
         C. Cartier, J. Hwang, and E. Copeland,
            {\it ib. id.} {\bf 64} 103504 (2001), (astro-ph/0106197).
\bibitem{Rab}
         H. Noh and J. Hwang, Phys. Rev. D {\bf 55}, 5222 (1997);
            {\it ib. id.} {\bf 59} 047501 (1999).
\end{thebibliography}
\end{document}